\documentclass[10pt,twocolumn,letterpaper]{article}

\usepackage[pagenumbers]{cvpr} 
\definecolor{cvprblue}{rgb}{0.21,0.49,0.74}
\usepackage{pifont}
\usepackage[pagebackref,breaklinks,colorlinks,allcolors=cvprblue]{hyperref}
\usepackage{xcolor} 
\usepackage{arydshln}
\usepackage{amsmath}     
\usepackage{amssymb}
\usepackage{tikz}   
\usepackage{multirow} 
\usepackage{tcolorbox}
\usepackage{array}
\usepackage{graphicx}
\usepackage{multirow}
\usepackage{booktabs}
\usepackage{adjustbox}
\usepackage{colortbl}
\usepackage{bm}
\definecolor{cvprblue}{rgb}{0.21,0.49,0.74}
\definecolor{red_lc}{RGB}{132, 27, 2}
\definecolor{blue_lc}{RGB}{22, 57, 111}
\definecolor{grey_lc}{RGB}{154, 152, 153}

\definecolor{marimba1}{RGB}{136, 20, 0}
\definecolor{marimba2}{RGB}{200, 105, 100}
\definecolor{cat1}{RGB}{20, 59, 104}
\definecolor{cat2}{RGB}{54, 116, 170}
\definecolor{cat3}{RGB}{88, 142, 174}
\definecolor{wolf1}{RGB}{220, 96, 31}
\definecolor{wolf2}{RGB}{254, 189, 66}
\definecolor{green_lc}{RGB}{93, 174, 86}
\definecolor{gray_lc2}{RGB}{89, 89, 89}
\definecolor{red_deep_lc}{RGB}{192, 0, 0}
\definecolor{blue_deep_lc}{RGB}{0, 32, 96}
\def\paperID{8210} 
\def\confName{CVPR}
\def\confYear{2025}

\title{Dynamic Derivation and Elimination: Audio Visual Segmentation with Enhanced Audio Semantics}

\author{Chen Liu$^{1,4}$, Liying Yang$^{5}$, Peike Li$^{3}$,  Dadong Wang$^{4}$, Lincheng Li$^{2}$,  Xin Yu$^{1}$\footnotemark[1]\\
$^{1}$ The University of Queensland, 
$^{2}$ NetEase Fuxi AI Lab,
$^{3}$ Matrix Verse AI, \\
$^{4}$ CSIRO Data61,
$^{5}$ Macau University of Science and Technology\\
{\tt\small\ yenanliu36@gmail.com, \tt\small\ xin.yu@uq.edu.au} 
}

\begin{document}
\maketitle
\renewcommand{\thefootnote}{\fnsymbol{footnote}}
\footnotetext[1]{Corresponding author.}
\begin{abstract}
Sound-guided object segmentation has drawn considerable attention for its potential to enhance multimodal perception.
Previous methods primarily focus on developing advanced architectures to facilitate effective audio-visual interactions, without fully addressing the inherent challenges posed by audio natures, \emph{\ie}, (1) feature confusion due to the overlapping nature of audio signals, and (2) audio-visual matching difficulty from the varied sounds produced by the same object.
To address these challenges, we propose Dynamic Derivation and Elimination (DDESeg): a novel audio-visual segmentation framework.
Specifically, to mitigate feature confusion, DDESeg reconstructs the semantic content of the mixed audio signal by enriching the distinct semantic information of each individual source, deriving representations that preserve the unique characteristics of each sound.
To reduce the matching difficulty, we introduce a discriminative feature learning module, which enhances the semantic distinctiveness of generated audio representations.
Considering that not all derived audio representations directly correspond to visual features (e.g., off-screen sounds), we propose a dynamic elimination module to filter out non-matching elements.
This module facilitates targeted interaction between sounding regions and relevant audio semantics.
By scoring the interacted features, we identify and filter out irrelevant audio information, ensuring accurate audio-visual alignment.
Comprehensive experiments demonstrate that our framework achieves superior performance in AVS datasets.
Our code is \href{https://github.com/YenanLiu/DDESeg}{here}.
\end{abstract}
\vspace{-2.0em}

\begin{figure}[t]
\begin{center}
\vspace{1em}
\includegraphics[width=0.9\linewidth]{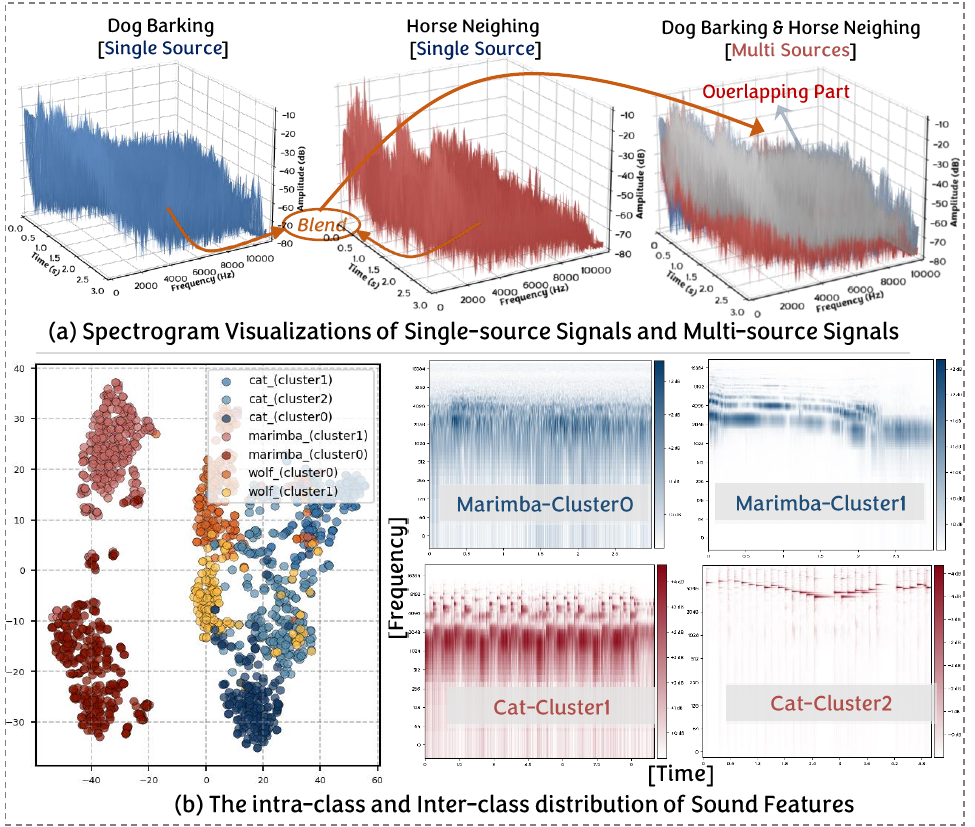}
\end{center}
\vspace{-1.5em}
\caption{
The illustration of \textit{\textbf{\textcolor{red_lc}{feature confusion}}} and \textit{\textbf{\textcolor{blue_lc}{audio-visual matching difficulty}}}.
(a) \textit{\textbf{\textcolor{red_lc}{Feature confusion}}} denotes the challenge of distinguishing or separating individual sound sources in a mixed audio signal, especially with overlapping frequency, timbre, or spatial cues, which impedes accurate semantic extraction. 
(b) Significant amplitude and frequency variations within sounds produced by the same object lead to large intra-class variation, introducing \textit{\textbf{\textcolor{blue_lc}{audio-visual matching difficulty}}}. This complicates the model's ability to align audio and visual modalities.
}
\label{fig:teaser}
\end{figure}

\begin{figure*}[htp]
\begin{center}
\includegraphics[width=0.9\linewidth]{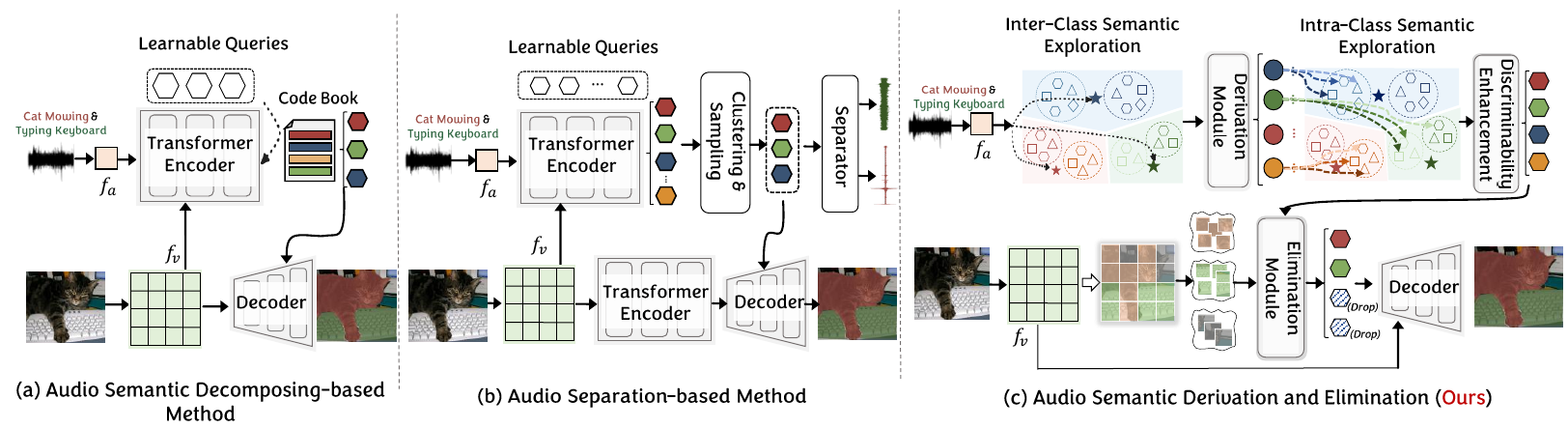}
\end{center}
\vspace{-2em}
\caption{
Three methods for achieving precise audio-visual alignment:
(a) \textbf{Audio Semantic Decomposition \cite{li2024qdformer}:} Models multi-source semantic space as a Cartesian product of single-source subspaces, employing product quantization and a shared codebook to decompose audio features into compact semantic tokens.
(b) \textbf{Audio Separation \cite{chen2024cpm}:} Devising a branch to decode audio-visual fused features into separated audio signals;
(c) \textbf{Audio Semantic Derivation and Elimination (Ours):} Derives distinct semantic representations for each source from a mixed audio signal by exploring inter-class relationships. Furthermore, derived semantic representations are refined through intra-class relationships, while irrelevant audio representations are excluded under visual guidance.
}
\label{fig:task_compare}
\vspace{-1.0em}
\end{figure*}

\section{Introduction}
\label{sec:intro}
Audio-visual segmentation (AVS) aims to localize sound-emitting objects by generating either a binary map of audible regions or a semantic map categorizing the objects \cite{zhou2024audio, zhou2022audio}. 
Compared to other modality-guided segmentation tasks (\emph{e.g.} text or visual) \cite{ding2021vision, ding2023mevis, wang2024prompting, wu2023referring}, audio-driven segmentation presents unique challenges due to the inherent overlap and intra-class variability of sounds \cite{zhao2018sound, owens2018audio, zhao2019sound, arandjelovic2018objects}.
While previous methods have focused on enhancing audio-visual interactions through learnable audio queries \cite{liu2023audio_a, huang2023discovering, wang2024prompting} or incorporating text \cite{li2022learning, li2023progressive, li2024object} into audio semantics, they usually overlook the complexities introduced by audio's unique properties. 
In this work, we propose solutions specifically tailored to address the challenges that audio's inherent characteristics pose to AVS.

We identify two key challenges posed by audio signals.
The first is \textit{\textbf{feature confusion due to overlapping audio signals}}, where simultaneous sounds combine across time and frequency domains, forming a complex composite signal.
As illustrated in Fig. \ref{fig:teaser} (a), the signals of \textit{\textbf{\textcolor{red_lc}{horse neighing}}} and \textit{\textbf{\textcolor{blue_lc}{dog barking}}} exhibit significant overlap (\textit{\textbf{\textcolor{grey_lc}{grey region}}}) in frequency and amplitude, complicating the isolation of unique features for each source. 
Consequently, extracted audio semantics may fail to capture the distinct semantic characteristics of each sound, leading to unreliable audio-visual alignment.
Recent methods \cite{li2024qdformer, chen2024cpm} assume that sound events occur independently and attempt to decompose or separate audio semantics from audio-visual fused features, as illustrated in Fig. \ref{fig:task_compare}.
However, different sources usually share frequencies within the spectrum, which poses challenges for these models to assign each signal component to the correct source accurately.
This overlap leads to information loss, with key semantic details either diluted or missing, resulting in incomplete and unreliable representations.

The second challenge is \textit{\textbf{audio-visual matching difficulty due to intra-class variations of sounds}}. 
As depcited in Fig. \ref{fig:teaser} (b), sounds from the cat (\tikz \filldraw[fill=cat1] (0,0) circle (0.5ex); \tikz \filldraw[fill=cat2] (0,0) circle (0.5ex); \tikz \filldraw[fill=cat3] (0,0) circle (0.5ex);), marimba (\tikz \filldraw[fill=marimba1, draw=black] (0,0) circle (0.5ex); \tikz \filldraw[fill=marimba2, draw=black] (0,0) circle (0.5ex);), and wolf (\tikz \filldraw[fill=wolf1, draw=black] (0,0) circle (0.5ex); \tikz \filldraw[fill=wolf2, draw=black] (0,0) circle (0.5ex);) categories in the AVS dataset exhibit significant intra-class variation.\footnote{The features are extracted by the state-of-the-art audio classification model HTSAT \cite{chen2022hts}, fine-tuned on the AVS dataset \cite{zhou2024audio}.}
These variations primarily arise from considerable differences in the frequency and temporal characteristics of sounds produced by the same object \cite{arandjelovic2018objects, owens2018audio}. For example, cat sounds can be further classified into subcategories such as ``\texttt{cat caterwauling}", ``\texttt{cat growling}", ``\texttt{cat hissing}", and ``\texttt{cat meowing}". 
The varied sound produced by the same visual category object poses challenges for the model to consistently associate these diverse sounds with the same object, thereby hindering accurate audio-visual alignment.

In this paper, we present a novel AVS framework, Dynamic Derivation and Elimination (\textbf{DDESeg}), to address these challenges. 
Rather than separating or decoupling the audio semantics from the mixed one, DDESeg's Dynamic Derivation Module first identifies the possible sound semantic representations within the input audio signal.
By supplementing incomplete information related to independent sound sources within the mixed audio signal, we derive multiple distinct representations, each conveying independent semantic information, as depicted in Fig. \ref{fig:task_compare} (c).
Once we obtain these derived audio semantic representations, we further enhance their discriminability by exploring intra-class semantic relationships.
Inspired by the Generalized Laplace Operator \cite{yu2016unitbox, belkin2003laplacian}, our method captures feature distinctions by calculating differences and adaptively aggregating them. 
We then scale the learned discriminative features to refine the derived semantics. 
In this fashion, we enhance the discriminability of derived semantics while preserving the integrity of the original semantic space, effectively mitigating misalignment issues caused by the intra-class variation.

Furthermore, considering the sounds may not correspond to visual regions in the image (\emph{e.g.,} off-screen sounds), we introduce the Dynamic Elimination Module (DEM) to eliminate ineffective audio semantics.
Specifically, as shown in Fig. \ref{fig:task_compare} (c), we first extract key semantics from the visual frame to isolate sound features associated with specific objects. 
The highly concentrated visual and audio semantic representations interact through an attention-based layer to generate the fused features.
Afterward, we assess the relevance score of these fused features and utilize these scores to identify and weaken irrelevant audio semantics.
By selectively emphasizing relevant audio features, we minimize interference from unrelated audio signals, ensuring robust audio-visual alignment.

\noindent In summary, our contributions are summarized as follows:

\begin{itemize}
\item We present an effective audio-visual segmentation framework to address the unique challenges in AVS arising from audio characteristics.
\item We propose the DDM, which first derives multiple audio representations from the input audio signal to address the feature fusion problem, and then enhances these derived semantics by devising discriminative features to enhance the audio-visual alignment.
\item We develop DEM to eliminate derived audio representations that lack a matching visual region in the frame, enabling precise audio-visual alignment.
\item Extensive experiments demonstrate that our method achieves competitive segmentation performance, significantly outperforming previous state-of-the-art methods across all AVS benchmarks, \emph{e.g.}, a 6.4\% improvement in $\mathcal{J}\&\mathcal{F}_\beta$ over the second-best method on AVS-Semantic.
\end{itemize}

\section{Related Work}
\label{sec:formatting}


\noindent\textbf{Referring Image Segmentation.} Referring Image Segmentation focuses on identifying and segmenting target objects within an image based on textual descriptions \cite{wu2024towards, chng2024mask, liu2024referring, xu2023meta, hui2023language, wu2023referring, hu2023beyond, ding2021vision}.
The self-attention mechanism \cite{vaswani2017attention} in transformers has inspired many studies that utilize cross-attention blocks to enhance cross-modal alignment \cite{wang2022cris, kim2022restr, luo2024soc}. 
For instance, VLT \cite{ding2021vision} utilizes a cross-attention module to generate query vectors that capture multimodal features, which are then employed to query the image through the transformer decoder. 
Similarly, DMMI \cite{hu2023beyond} introduces dual decoder branches to facilitate bidirectional information flow, enhancing image-text interactions.

Query-based transformer architectures \cite{zou2023generalized, miao2023spectrum, zou2024segment, ding2023mevis} have gained popularity in multimodal-guided segmentation tasks due to their ability to handle complex interactions between different data types, such as images and text.
Models like X-Decoder \cite{zou2023generalized}, SEEM \cite{zou2024segment}, and QaP \cite{zou2024segment} employ text tokens as queries, a design that has demonstrated strong performance. 
Inspired by these successes, some AVS architectures have adopted query-based transformers, utilizing audio tokens as queries \cite{liu2024bavs, liu2023audio, ma2024stepping}. 
However, the unique characteristics of audio, such as its overlapping and temporal nature, necessitate more specialized considerations when designing audio-visual alignment modules.

\noindent\textbf{Audio Visual Segmentation.}
Given an audio signal, audio-visual segmentation aims to segment and identify sound-emitting objects in images or videos \cite{zhou2024audio, sun2024unveiling, chen2024cpm, zhou2022audio, chen2024unraveling, mao2023contrastive, liu2023audio_a, mao2023multimodal, liu2024annotation, huang2023discovering, gao2024avsegformer, hao2024improving, wang2024prompting, yan2024referred, yang2024cooperation, li2023catr}. 
Unlike images or text, which provide discrete and localized information, audio inherently introduces ambiguity due to its overlapping nature, complicating the alignment between audio cues and visual regions. 
Recent studies \cite{liu2023audio, liu2024bavs, li2024qdformer, chen2024cpm} address this challenge by employing sound separation techniques to decompose mixed audio into individual sound components, or by leveraging audio classification results to identify relevant audio semantics. 
For instance, \citet{liu2024bavs} and \citet{sun2024unveiling} extract semantics from mixed audio by first obtaining classification results and then applying a threshold to filter out irrelevant items from the multi-class outputs.
QDFormer \cite{li2024qdformer} assumes sound events are independent and utilizes a quantization-based decomposition method to separate them, while CPM \cite{chen2024cpm} decodes audio-visual queries into spectrograms and applies an separation loss for supervision.

Despite these advancements, the challenges posed by the unique characteristics of audio remain unresolved. 
To address this, our method derives distinct, single-sounding source semantic representations from the input audio signal and explores intra-class relationships to enhance their discriminability.


\begin{figure*}[htp]
\begin{center}
\includegraphics[width=0.95\linewidth]{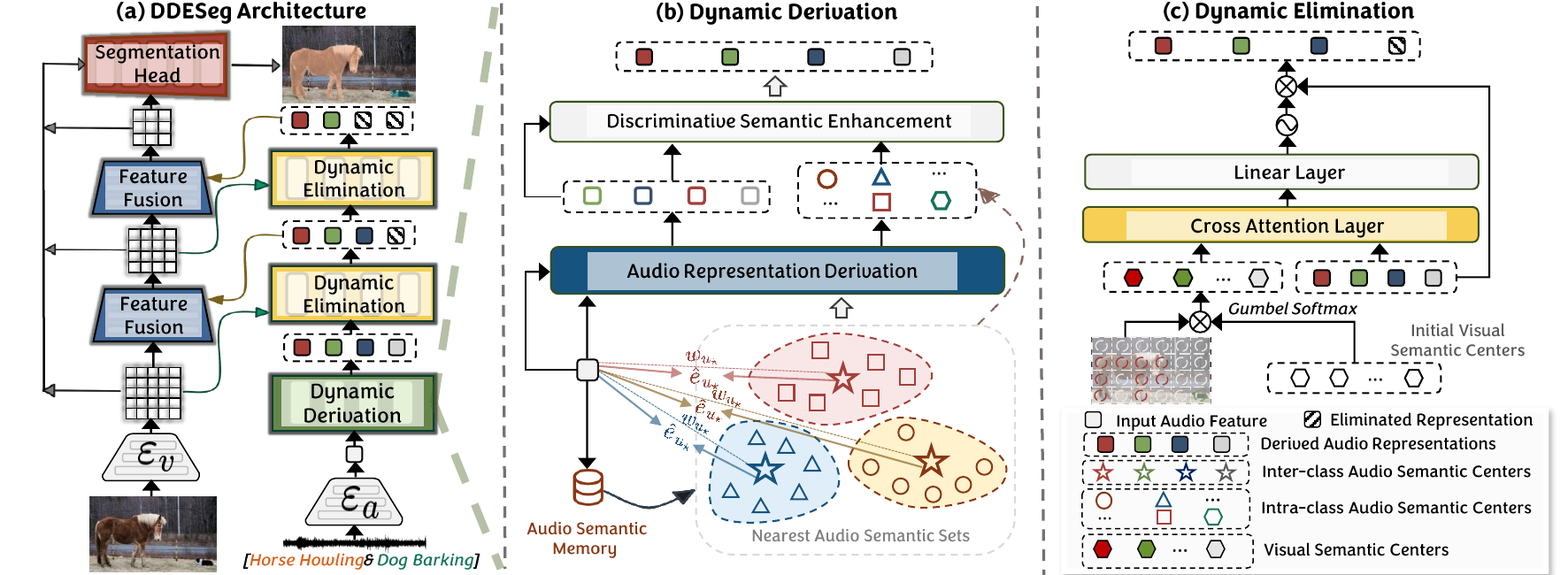}
\end{center}
\vspace{-1.5em}
\caption{
Overview of DDESeg architecture and its key components:
(a) \textbf{Framework Architecture.} Overview of our dual-branch framework that hierarchically processes and aligns audio-visual features. Through progressive multi-stage alignment, the framework fuses cross-modal information and generates precise pixel-wise segmentation maps via the segmentation head.
(b) \textbf{Dynamic Derivation Module.} This module generates multiple audio representations from the input audio feature and explores intra-class relationships to equip discriminative features for each derived representation.
(c) \textbf{Dynamic Elimination Module.} DEM eliminates audio representations that do not correspond to visual regions by evaluating the relevance between audio representations and learned image semantic representations.
}
\label{fig:pipeline}
\vspace{-1.5em}
\end{figure*}

\section{Proposed Method}
We propose \textbf{DDESeg}, a dynamic audio-visual segmentation framework that achieves precise audio-visual alignment through multi-stage cross-modal interactions (\emph{cf}. Fig. \ref{fig:pipeline} (a)). 
The framework consists of three key components: a dynamic derivation module (\S\ref{Sec:DDM}) that derives single-source audio semantic representations from the input audio signal by exploring inter- and intra-class audio semantic relationships, a dynamic elimination module (\S\ref{Sec:DDE}) that filters out audio representations irrelevant to the visual content, and a feature fusion module (\S\ref{Sec:FF}) that progressively aligns the refined audio features with visual features. 

The workflow of our framework is mathematically expressed as follows:
\begin{equation}
\begin{aligned}
    \hat{M} &= \mathcal{D} \left( \left\{ \mathcal{F}_f \left(   
    \hat{A}^{l}, V^{l} \right) \right\}_{l=1}^{L} \right), \\
    \hat{A}^{(l)} &= 
    \begin{cases} 
    \Phi_D(F_a), & \text{if } l = 1, \\
    \Phi_E(\hat{A}^{(l-1)}, V^{l}), & \text{if } l > 1.
    \end{cases}
\end{aligned}
\end{equation}
Here, $\hat{M} \in [0, C]^{H \times W}$ denotes the predicted segmentation mask for the input audio-visual pair, and $L$ represents the number of stages in our framework. 
The modules $\Phi_D$, $\Phi_E$, $\mathcal{F}_f$, and $\mathcal{D}$ correspond to the audio semantic derivation, audio semantic elimination, feature fusion, and mask decoding operations, respectively. 
$\hat{A}^l$ represents the derived audio semantic representations at stage $l$. The visual features $V^l \in \mathbb{R}^{H_l \times W_l \times D}$ are either extracted by the visual backbone $\mathcal{E}_V$ (when $l=1$) or obtained from the $l$-th feature fusion layer (when $l>1$), where $H_l \times W_l$ denotes the spatial resolution and $D$ is the channel dimension.
$F_a$ represents the audio feature extracted from the input audio signal.
We employ the standard segmentation loss for training supervision, as detailed in \S \ref{sec:loss}.


\subsection{Dynamic Derivation Module}
\label{Sec:DDM}
The Dynamic Derivation Module (DDM) generates audio semantic representations with distinct single-source semantics based on the input audio signal and semantic memory (\emph{cf.} Fig. \ref{fig:pipeline} (b)).
This process involves three key steps: \textbf{Step1}: Constructing a robust audio semantic memory from single-sounding source signals to capture clean and distinct semantic features.
\textbf{Step2}: Retrieving the $K$ nearest semantic centers of the input audio signal and using them to transform the input audio features into new representations with enhanced semantic distinctiveness;
and \textbf{Step3}: Learning intra-class feature differences and adaptively refining the derived audio representations to further enhance their discriminability.
 
\noindent\textbf{Step1: Semantic Memory Construction.}
To construct a robust semantic memory, we employ a pre-trained foundation model to extract audio features from single-source audio signals, providing a solid basis for capturing distinct and meaningful semantics.
For each class $c$, we analyze its feature distribution through hierarchical clustering, starting with the global centroid $\mu^c$, which is computed as: 
\begin{equation} \mu^c = \frac{1}{n_c} \sum_{i=1}^{n_c} x_i^c, 
\end{equation} 
where $x_i^c$ are the extracted audio features, and $n_c$ is the number of features in class $c$. 
This global centroid serves as a reference for deriving audio representations in Step 2.
 
To capture intra-class variations, we partition each class $c$ into $k$ clusters $\{C_j^c\}_{j=1}^m$ via the k-means algorithm, with sub-cluster centroids $\mu_j^c$ computed as: 
\begin{equation}
    \mu_j^c =\frac{1}{|C_j^c|}\sum_{x_i^c \in C_j^c} x_i^c.
\end{equation}

For each cluster $C_j^c$, we select $m$ feature vectors closet to the cluster centroid $\mu_j^c$ as the discriminative features:
\begin{equation}
    x_{rep_j}^c = arg \min_{x_i^c\in C_j^c, |X_j|=m}\sum_{x_i^c\in X_j}||x_i^c -\mu_j^c||,
\end{equation}
where $X_j \subseteq C_j^c$. 
The resulting discriminative features $x_{rep_j}^c \in \mathbb{R}^{m \times d}$ form a rich semantic memory that will be leveraged in Step 3 to enhance the derived audio prototypes.

\noindent\textbf{Step2: Audio Prototype Derivation.}
The goal of this step is to derive multiple audio representations with unique audio semantics from the input audio feature $F_a$.
Given $F_a$ and the audio semantic memory, we first compute the Euclidean distance $d_c$ to each class center $\mu^c$, and identify the $K$ nearest audio semantic centers $\{\mu_i\}_{i=1}^K$.

Inspired by the Generalized Laplace Operator, which emphasizes distinctions between the features of a central point and its neighboring points to accurately measure and adjust for semantic differences, we first compute the edge feature $e_{u_i} \in \mathbb{R}^d$ for each selected nearest class center $\mu_i$. This edge feature captures the relationship between $F_a$ and $u_i$, expressed by:
\begin{equation}
    e_{u_i} = \phi_{\text{GELU}}(W_e(u_i - F_a) + b_e),
\end{equation}
where $W_e \in \mathbb{R}^{d \times d}$ and $b_e \in \mathbb{R}^d$ are learnable parameters. $\phi_{GELU}$ is the Gaussian error linear unit activation function.
Then, we compute a fusion weight $w_{u_i}$ for each edge feature to quantify its relevance and further attain the weighted edge features $\hat{e}_{u_i} \in \mathbb{R}^d$.
The process is defined as:
\begin{equation}
    w_{u_i} = \texttt{softmax}(W_{f1}e_{u_i}), \\
    \hat{e}_{u_i} = w_{u_i}e_{u_i},
\end{equation}
where $W_{f1} \in \mathbb{R}^{1 \times d}$ is learnable weight matrix.

Based on the weighted edge features $\hat{e}_{u_i}$, we compute the compensation $\Delta a_{u_i} \in \mathbb{R}^d$ to bridge the semantic discrepancy between the input audio feature $F_a$ and class centers.
This is mathematically formulated as:
\begin{equation}
    \Delta a_{u_i} = \phi_{\text{tanh}}(W_{o1}(F_a + \hat{e}_{u_i}) + b_{o1}),
\end{equation}
where $W_{o1} \in \mathbb{R}^{d \times d}$ and $b_{o1} \in \mathbb{R}^{d}$ are learnable parameters.
The derived audio representation $a_i$ is then given by:
\begin{equation}
   a_i = F_a + \Delta a_{u_i}.
\end{equation}
Leveraging this design, we generate $K$ distinct audio representations $A=\{a_i\}_{i=1}^K \in \mathbb{R}^{K \times d}$ from the input feature $F_a$ through precise, semantically coherent adjustments. 

\noindent\textbf{Step3: Discriminative Semantic Enhancement.}
In this step, our goal is to capture the subtle differences between derived audio semantics and discriminative intra-class audio features to mitigate semantic bias arising from large intra-class variations.
Specifically, for each derived audio representation $a_i$, we compute the edge features $e_{c_i,j} \in \mathbb{R}^{m \times d}$ between $a_i$ and the discriminative intra-class feature $x^c_{rep_j}$, defined as follows:
\begin{equation} e_{c_i,j} = \phi_{\text{GELU}}(W_d(x^c_{rep_j} - a_i) + b_d), \end{equation}
where $W_d \in \mathbb{R}^{d \times d}$ and $b_d \in \mathbb{R}^d$ are learnable parameters.

Next, we calculate the fusion weights for the intra-class edge features, resulting in edge weights $w_{c_i,j} \in \mathbb{R}^{1 \times m}$ and the corresponding weighted edge features $\hat{e}_{c_i,j} \in \mathbb{R}^{1 \times d}$, expressed as:
\begin{equation} w_{c_i,j} = \texttt{softmax}(W_{f2}e_{c_i,j}), \quad \hat{e}_{c_i,j} = w_{c_i,j} \cdot e_{c_i,j}, 
\end{equation}
where $W_{f2} \in \mathbb{R}^{1 \times d}$ is a learnable weight matrix.

Afterwards, we compute the offset vectors $\Delta a_{c_i,j} \in \mathbb{R}^{1 \times d}$ for intra-class refinement, defined as:
\begin{equation} 
\Delta a_{c_i,j} = \phi_{\text{tanh}}(W_{o2}(\hat{e}_{c_i,j}) + b_{o2}), 
\end{equation}
where $W_{o2} \in \mathbb{R}^{d \times d}$ and $b_{o2} \in \mathbb{R}^{d}$ are learnable parameters.
Finally, we enhance each derived audio representation by incorporating the learned offset.
Unlike the additive adjustment applied in Step 2, this step refines $\hat{a}_{i}$ by scaling its features to emphasize discriminative aspects without deviating from the original semantic space, as defined by:
\begin{equation} \hat{a}_{i} = a_i \odot (1 + \Delta a_{c_i,j}), \end{equation}
where $\odot$ denotes element-wise multiplication, and $\hat{a}_{i} \in \mathbb{R}^{1 \times d}$ represents the refined audio semantics. 
After this process, we obtain the set of enhanced audio representations $\hat{A}=\{\hat{a}_i\}_{i=1}^K \in \mathbb{R}^{K \times d}$.
Notably, we employ addition to adjust inter-class semantic gaps, as it facilitates shifts between categories. 
In contrast, we apply scaling for intra-class enhancement, as it preserves the adjusted features within the same class. 

\begin{figure}[t]
\begin{center}
\includegraphics[width=0.9\linewidth]{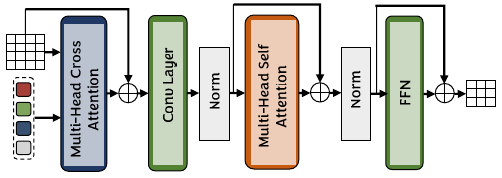}
\end{center}
\vspace{-2.0em}
\caption{Structure of the Feature Fusion Block in DDESeg. }
\label{fig:fusion_block}
\vspace{-1.5em}
\end{figure}

\begin{table*}[htp]
\caption{Quantitative comparisons on the AVS-Objects and AVS-Semantic datasets (see analysis details in \S \ref{sec:qualitative_ana}). `Trans.': Transformer-based architecture. Best results in \textbf{Bold}, while second best \underline{underlined}.}
\vspace{-0.5em}
\label{tab:avss}
\centering
\footnotesize
\setlength{\tabcolsep}{1.6mm}
\renewcommand\arraystretch{0.9}
\begin{tabular}{r|c|cccc|ccc|ccc}
\toprule
\multicolumn{1}{c|}{}                         & \multicolumn{1}{c|}{}                                  & \multicolumn{1}{c|}{}                                 & \multicolumn{3}{c|}{\textit{AVS-Objects-S4}}                                             & \multicolumn{3}{c|}{\textit{AVS-Objects-MS3}}                                            & \multicolumn{3}{c}{\textit{AVS-Semantic}}                                                 \\ \cmidrule{4-12} 
\multicolumn{1}{c|}{\multirow{-2}{*}{Methods}} & \multicolumn{1}{c|}{\multirow{-2}{*}{\shortstack{Visual \\ Backbone}}} & \multicolumn{1}{c|}{\multirow{-2}{*}{\shortstack{Audio \\ Backbone}}} & \multicolumn{1}{c}{$\mathcal{J}\&\mathcal{F}_m \uparrow$} & \multicolumn{1}{c}{$\mathcal{J} \uparrow$} & \multicolumn{1}{c|}{$\mathcal{F}_m \uparrow$}       & \multicolumn{1}{c}{$\mathcal{J}\&\mathcal{F}_m \uparrow$} & \multicolumn{1}{c}{$\mathcal{J} \uparrow$} & \multicolumn{1}{c|}{$\mathcal{F}_m \uparrow$}       & \multicolumn{1}{c}{$\mathcal{J}\&\mathcal{F}_m \uparrow$} & \multicolumn{1}{c}{$\mathcal{J} \uparrow$}  & \multicolumn{1}{c}{$\mathcal{F}_m \uparrow$}       \\ \midrule
TPAVI \cite{zhou2022audio}~\textcolor{gray}{\tiny{[ECCV22]}}                                         & PVT-V2-B5                                              & \multicolumn{1}{l|}{VGGish}                           & 83.3                     & 78.7                  & 87.9                         & 59.3                     & 54.0                  & 64.5                         & 32.5                     & 29.8                   & 35.2                         \\
AVSC \cite{liu2023audio}~\textcolor{gray}{\tiny{[ACM-MM23]}}                                         & Swin-Base                                              & \multicolumn{1}{l|}{VGGish}                           & 85.0                     & 81.3                  & 88.6                         & 62.6                     & 59.5                  & 65.8                         & \multicolumn{1}{c}{\_}   & \multicolumn{1}{c}{\_} & \multicolumn{1}{c}{\_}       \\
CATR \cite{li2023catr}~\textcolor{gray}{\tiny{[ACM-MM23]}}                                         & PVT-V2-B5                                              & \multicolumn{1}{l|}{VGGish}                           & 87.9                     & 84.4                  & 91.3                         & 68.6                     & 62.7                  & 74.5                         & 35.7                     & 32.8                   & 38.5                         \\
DiffusionAVS \cite{mao2023contrastive}~\textcolor{gray}{\tiny{[Arxiv23]}}                                  & PVT-V2-B5                                              & \multicolumn{1}{l|}{VGGish}                           & 85.7                     & 81.4                  & 90.0                         & 64.6                     & 58.2                  & 70.9                         & \multicolumn{1}{c}{\_}   & \multicolumn{1}{c}{\_} & \multicolumn{1}{c}{\_}       \\
ECMVAE \cite{mao2023multimodal}~\textcolor{gray}{\tiny{[ICCV23]}}                                       & PVT-V2-B5                                              & \multicolumn{1}{l|}{VGGish}                           & 85.9                     & 81.7                  & 90.1                         & 64.3                     & 57.8                  & 70.8                         & \multicolumn{1}{c}{\_}   & \multicolumn{1}{c}{\_} & \multicolumn{1}{c}{\_}       \\
AuTR \cite{liu2023audio_a}~\textcolor{gray}{\tiny{[Arxiv23]}}                                          & PVT-V2-B5                                              & \multicolumn{1}{l|}{VGGish}                           & 82.1                     & 77.6                  & 86.5                         & 72.0                     & 66.2                  & 77.7                         & \multicolumn{1}{c}{\_}   & \multicolumn{1}{c}{\_} & \multicolumn{1}{c}{\_}       \\
AQFormer \cite{huang2023discovering}~\textcolor{gray}{\tiny{[IJCAI23]}}                                      & PVT-V2-B5                                              & \multicolumn{1}{l|}{VGGish}                           & 85.5                     & 81.6                  & 89.4                         & 67.5                     & 62.2                  & 72.7                         & \multicolumn{1}{c}{\_}   & \multicolumn{1}{c}{\_} & \multicolumn{1}{c}{\_}       \\
AVSegFormer \cite{gao2024avsegformer}~\textcolor{gray}{\tiny{[AAAI24]}}                                  & PVT-V2-B5                                              & \multicolumn{1}{l|}{VGGish}                           & 86.8                     & 83.1                  & 90.5 & 67.2                     & 61.3                  & 73.0 & 40.1                     & 37.3                   & 42.8 \\
AVSBG \cite{hao2024improving}~\textcolor{gray}{\tiny{[AAAI24]}}                                        & PVT-V2-B5                                              & \multicolumn{1}{l|}{VGGish}                           & 86.1                     & 81.7                  & 90.4                         & 61.0                     & 55.1                  & 66.8                         & \multicolumn{1}{c}{\_}   & \multicolumn{1}{c}{\_} & \multicolumn{1}{c}{\_}       \\
GAVS \cite{wang2024prompting}~\textcolor{gray}{\tiny{[AAAI24]}}                                         & ViT-Base                                               & \multicolumn{1}{l|}{VGGish}                           & 85.1                     & 80.1                  & 90.0                         & 70.6                     & 63.7                  & 77.4                         & \multicolumn{1}{c}{\_}   & \multicolumn{1}{c}{\_} & \multicolumn{1}{c}{\_}       \\
BAVS \cite{liu2024bavs}~\textcolor{gray}{\tiny{[TMM24]}}                                         & Swin-Base                                              & \multicolumn{1}{l|}{Beats}                            & 86.2                     & 82.7                  & 89.8                         & 62.8                    & 59.6                  & 65.9                         & 35.6                     & 33.6                   & 37.5                         \\
COMBO \cite{yang2024cooperation}~\textcolor{gray}{\tiny{[CVPR24]}}                                         & PVT-V2-B5                                              & \multicolumn{1}{l|}{VGGish}                           & 88.3                     & 84.7                  & 91.9                         & 65.2                     & 59.2                  & 71.2                         & 44.1                     & 42.1                   & 46.1                         \\
QDFormer \cite{li2024qdformer}~\textcolor{gray}{\tiny{[CVPR24]}}                                      & Swin-Tiny                                              & \multicolumn{1}{l|}{VGGish}                           & 83.9                     & 79.5                  & 88.2                         & 64.0                     & 61.9                  & 66.1                         & \multicolumn{1}{c}{\_}   & 53.4                   & \multicolumn{1}{c}{\_}       \\
CAVP \cite{chen2024unraveling}~\textcolor{gray}{\tiny{[CVPR24]}}                                          & PVT-V2-B5                                              & \multicolumn{1}{l|}{VGGish}                           & \underline{90.5}                     & \underline{87.3}                 & \underline{93.6}                         & 72.7                     & \underline{67.3}                  & 78.1                         & 55.3                     & 48.6                   & 62.0                         \\
AVSStone\footnote{}  \cite{ma2024stepping}~\textcolor{gray}{\tiny{[ECCV24]}}                                     & Swin-Base                                              & \multicolumn{1}{l|}{VGGish}                           & 87.3                     & 83.2                  & 91.3                         & 72.5                     & \underline{67.3}                  & 77.6                         & \underline{61.5}                    & \underline{56.8}                   & \underline{66.2}                         \\
BiasAVS  \cite{sun2024unveiling}~\textcolor{gray}{\tiny{[ACM-MM24]}}                             & Swin-Base                                              & \multicolumn{1}{l|}{VGGish}                           & 88.2                     & 83.3                  & 93.0                         & \underline{74.0}                     & 67.2                  & \underline{80.8}                         & 47.2                     & 44.4                   & 49.9                         \\ \midrule
\rowcolor[HTML]{e2f0d9} 
\textcolor{gray_lc2}{\textbf{\textsc{DDESeg (Ours)}}}                                          & \textcolor{gray_lc2}{PVT-V2-B5}                                                 & \multicolumn{1}{l|}{\textcolor{gray_lc2}{VGGish}}                           & \textcolor{gray_lc2}{92.0}                     & \textcolor{gray_lc2}{90.3}                  & \textcolor{gray_lc2}{92.2}                         & \textcolor{gray_lc2}{74.7}                     & \textcolor{gray_lc2}{69.7}                  & \textcolor{gray_lc2}{79.6}                         & \textcolor{gray_lc2}{65.7}                     & \textcolor{gray_lc2}{60.8}                   & \textcolor{gray_lc2}{70.6}                         \\
\rowcolor[HTML]{e2f0d9} 
\textcolor{gray_lc2}{\textbf{\textsc{DDESeg (Ours)}}}                                           & \textcolor{gray_lc2}{Trans.}                                                  & \multicolumn{1}{l|}{\textcolor{gray_lc2}{VGGish}}                           & \textcolor{gray_lc2}{92.7}                     & \textcolor{gray_lc2}{91.8}        &  \textcolor{gray_lc2}{93.5}               &  \textcolor{gray_lc2}{75.6}         &  \textcolor{gray_lc2}{70.1}         &  \textcolor{gray_lc2}{81.0}            & \textcolor{gray_lc2}{67.0}         & \textcolor{gray_lc2}{62.3}        & \textcolor{gray_lc2}{71.6}                \\
\rowcolor[HTML]{a9d18e} 
\textbf{\textsc{DDESeg (Ours)}}                                           & Trans.                                                  & \multicolumn{1}{l|}{HTSAT}                                                 & \textbf{94.2}                     & \textbf{92.4}         & \textbf{95.9}                & \textbf{77.9}            & \textbf{72.3}       & \textbf{83.4}                & \textbf{67.9}            & \textbf{63.4}          & \textbf{72.3}               \\ \bottomrule
\end{tabular}
\vspace{-2.0em}
\end{table*}

\subsection{Dynamic Elimination Module}
\label{Sec:DDE}
The Dynamic Elimination Module addresses a key challenge in audio-visual segmentation: ensuring that derived audio representations correspond to actual sound-emitting objects in the visual scene. 
Given derived audio representation $\hat{A} \in \mathbb{R}^{K \times d}$ and visual features $V \in \mathbb{R}^{H \times W \times d}$, this module leverages visual guidance to filter out irrelevant audio tokens, enhancing the discriminability of audio-visual associations (\emph{cf.} Fig. \ref{fig:pipeline} (c)).
This filtering step is crucial for achieving consistent audio-visual alignment.

\noindent\textbf{Soft Clustering of Visual Features.}
To obtain compact yet representative visual features, we employ a soft clustering approach with $K$ visual semantic centers $C_v \in \mathbb{R}^{K \times d}$. The assignment of visual features to these centers is computed through a Gumbel-Softmax operation:
\begin{equation}
    O = \frac{\text{exp}((VC_v^T) + g)/\tau}{\sum_{k=1}^N\text{exp}((VC_v^T) + g) / \tau},
\end{equation}
where $O \in \mathbb{R}^{HW \times K}$ represents the assignment matrix, $g$ consists of i.i.d. random samples from a \texttt{Gumbel(0, 1)} distribution, and $\tau$ serving as a temperature parameter.

The visual semantic centers are then refined through weighted aggregation:
\begin{equation}
    C_v = OC_v.
\end{equation}

\noindent\textbf{Semantic Scoring and Elimination.}
To evaluate the relevance of each audio token to the visual centers, we introduce a discriminative scoring mechanism. 
Specifically, we first perform cross-modal interaction between the audio features $\hat{A}$ and the visual centers $C_v$,  producing an audio-visual interacted feature $F_{av} \in \mathbb{R}^{k \times d}$.
This process is formulated as:
\begin{equation}
F_{av} = \texttt{MCA}(\hat{A}, C_v).
\end{equation}
where MCA denotes Multi-head Cross-Attention.

After that, we pass the interacted feature $F_{av}$ through a linear layer followed by a sigmoid activation function to obtain the discriminative score $S \in [0, 1]^k$ for each derived audio semantic, expressed by:
\begin{equation}
    S = \sigma(\texttt{MLP}(F_{av})).
\end{equation}
These scores serve as weights to weaken audio representations where no visual match can be found.
The updated process is formulated as follows:
\begin{equation}
    \hat{A} = S \cdot \hat{A}.
\end{equation}
This approach ensures that relevant audio semantics contribute more significantly to the audio-visual fused features.

\subsection{Feature Fusion}
\label{Sec:FF}
The Feature Fusion module performs hierarchical cross-modal interaction through multiple fusion blocks (\emph{cf}. Fig. \ref{fig:fusion_block}).
Each block employs a cross-attention architecture, with derived audio representations serving as queries and visual features as keys and values, enabling precise audio-visual alignment. 
After the cross-attention operation, a convolutional layer reduces spatial dimensions, distilling essential semantic information as in \cite{cheng2021maskformer, cheng2021mask2former}. 
The module concludes with a self-attention mechanism and feed-forward network to capture global contextual relationships, ensuring comprehensive feature integration across modalities.

\subsection{Loss Function}
\label{sec:loss}
Given the predicted segmentation map $\hat{m}$ and its corresponding ground truth $m$, we optimize our network using a weighted combination of three complementary loss terms:
\begin{equation}
    \begin{array}{cc}
         \mathcal{L}(m, \hat{m}) = \lambda_{dice}\mathcal{L}_{dice} + \lambda_{bce}\mathcal{L}_{bce} + \lambda_{iou}{L}_{iou},
    \end{array}
\end{equation}
where $\mathcal{L}_{dice}$ \cite{li2019dice} and $\mathcal{L}_{iou}$ \cite{yu2016unitbox} both optimize region-based overlap metrics from different perspectives, while $\mathcal{L}_{bce}$ provides pixel-wise supervision.
The balancing weights $\lambda_{dice}$, $\lambda_{bce}$, and $\lambda_{iou}$ are set to 5, 5, and 2, respectively.

\begin{figure*}[htp]
\begin{center}
\includegraphics[width=0.85\linewidth]{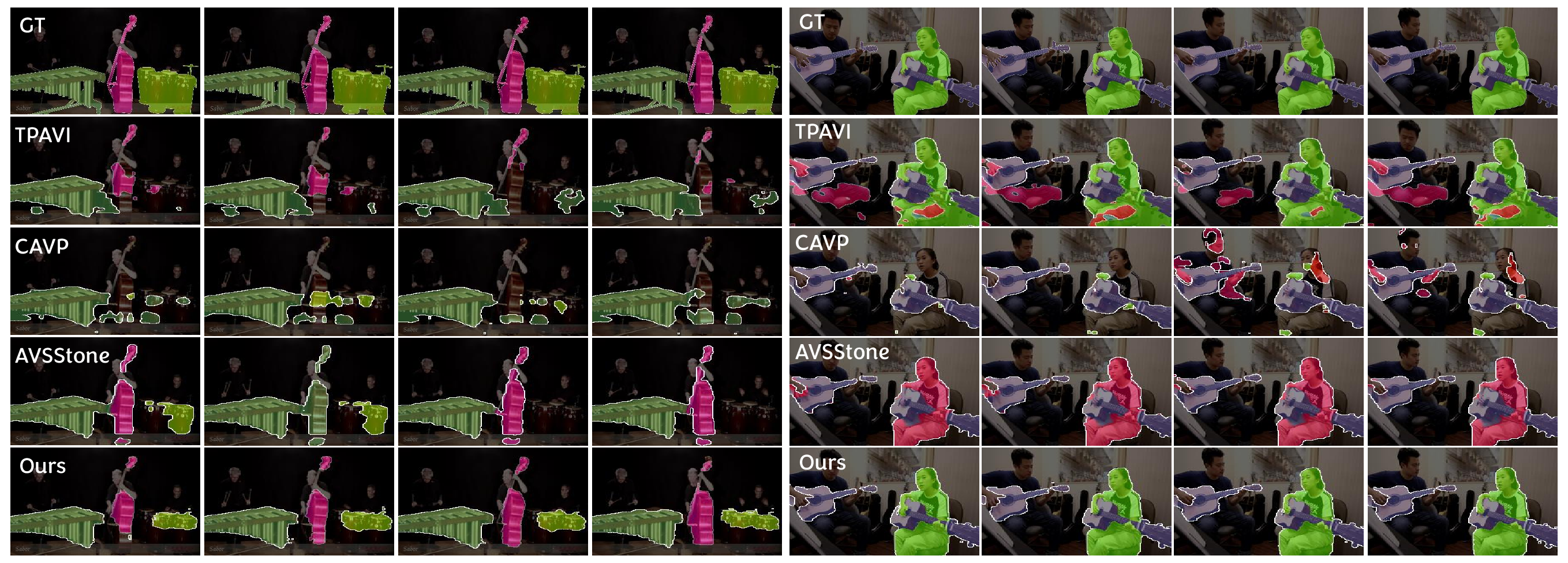}
\end{center}
\vspace{-2em}
\caption{Qualitative results on AVSBench-Semantic (\S \ref{sec:qualitative_ana}). Our method achieves precise localization in multi-source cases.}
\label{fig:com_avss}
\vspace{-1.0em}
\end{figure*}

\begin{table*}[htp]
\caption{Quantitative comparisons on the VPO datasets (\S \ref{sec:qualitative_ana}). `CPM*' indicates that training and testing are performed on images with the original resolution, while other methods employ images resized to 224 × 224 pixels. Best results in \textbf{Bold}, while second best \underline{underlined}.}
\vspace{-0.5em}
\label{tab:vpo}
\centering
\footnotesize
\setlength{\tabcolsep}{1.0mm}
\renewcommand\arraystretch{0.9}
\begin{tabular}{r|c|c|c|ccc|ccc|ccc}
\toprule
\multicolumn{1}{c|}{\multirow{2}{*}{Methods}} & \multicolumn{1}{c|}{\multirow{2}{*}{\shortstack{Visual \\ Backbone}}} & \multicolumn{1}{c|}{\multirow{2}{*}{\shortstack{Audio \\ Backbone}}} & \multicolumn{1}{c|}{\multirow{2}{*}{\shortstack{Trainable \\ Params}}} & \multicolumn{3}{c|}{\textit{VPO-SS}}                                                 & \multicolumn{3}{c|}{\textit{VPO-MS}}                                                 & \multicolumn{3}{c}{\textit{VPO-MSMI}}                                                             \\ \cmidrule{5-13} 
\multicolumn{1}{c|}{}                        & \multicolumn{1}{c|}{}                                 & \multicolumn{1}{c|}{}                                & \multicolumn{1}{c|}{}                                   & \multicolumn{1}{c}{$\mathcal{J}\&\mathcal{F}_\beta \uparrow$} & \multicolumn{1}{c}{$\mathcal{J} \uparrow$} & \multicolumn{1}{c|}{$\mathcal{F}_\beta \uparrow$} & \multicolumn{1}{c}{$\mathcal{J}\&\mathcal{F}_\beta \uparrow$} & \multicolumn{1}{c}{$\mathcal{J} \uparrow$} & \multicolumn{1}{c|}{$\mathcal{F}_\beta \uparrow$} & \multicolumn{1}{c}{$\mathcal{J}\&\mathcal{F}_\beta \uparrow$} & \multicolumn{1}{c}{$\mathcal{J} \uparrow$} & \multicolumn{1}{c}{$\mathcal{F}_\beta \uparrow$}             \\ \midrule

TPAVI  \cite{zhou2022audio}~\textcolor{gray}{\tiny{[ECCV22]}}                                       & PVT-V2-B5                                             & VGGish                                               & 101.32M                                                 & 44.63                     & 41.64                  & 47.62                  & 45.68                     & 42.3                   & 49.06                  & 43.19                     & 40.03                  & \multicolumn{1}{r}{46.34}          \\
AVSegFormer \cite{gao2024avsegformer}~\textcolor{gray}{\tiny{[AAAI24]}}                                 & PVT-V2-B5                                             & VGGish                                               & 186.05M                                                  & 45.94                     & 43.81                  & 48.06                  & 43.72                     & 47.3                   & 40.14                  & 49.93                     & 47.19                  & \multicolumn{1}{r}{52.67}          \\
CAVP \cite{chen2024unraveling}~\textcolor{gray}{\tiny{[CVPR24]}}                                        & ResNet50                                            & VGGish                                               & 119.79M                                                & 67.02                     & 58.81                  & 75.23                  & 61.32                     & 53.24                  & 69.39                  & 56.48                     & 48.18                  & \multicolumn{1}{r}{64.78}          \\
BiasAVS  \cite{sun2024unveiling}~\textcolor{gray}{\tiny{[ECCV24]}}                                   & SwinBase                                              & VGGish                                               & 107.12M                                                 & 67.46                     & 59.14                  & 75.78                  & 63.42                     & 55.61                  & 71.23                  & 57.94                     & 49.6                   & \multicolumn{1}{r}{66.27}          \\
AVSStone  \cite{ma2024stepping} ~\textcolor{gray}{\tiny{[ACM-MM24]}}                                    & SwinBase                                              & VGGish                                               & 114.63M                                                 & \underline{68.54}                     & \underline{59.72}                  & \underline{77.35}                  & \underline{64.26}                     & \underline{56.23}                  & \underline{72.29}                  & \underline{58.76}                     & \underline{50.11}                  & \multicolumn{1}{r}{\underline{67.40}}           \\ \cdashline{1-13}
CPM* \cite{chen2024cpm}~\textcolor{gray}{\tiny{[ECCV24]}}                                         & ResNet50                                              & VGGish                                               & \_                                                      & 73.49                     & 67.09                  & 79.88                  & 72.91                     & 65.91                  & 79.9                   & 68.07                     & 60.55                  & \multicolumn{1}{r}{75.58}          \\ \midrule
\rowcolor[HTML]{e2f0d9} 
\textcolor{gray_lc2}{\textbf{\textsc{DDESeg (Ours)}}}                                          & \textcolor{gray_lc2}{Trans.}                                                 & \textcolor{gray_lc2}{VGGish}                                               & \textcolor{gray_lc2}{103.84M}                                                 & \textcolor{gray_lc2}{73.78}                     & \textcolor{gray_lc2}{67.02}                  & \textcolor{gray_lc2}{80.54}                  & \textcolor{gray_lc2}{73.62}                     & \textcolor{gray_lc2}{67.01}                  & \textcolor{gray_lc2}{80.23}                  & \textcolor{gray_lc2}{67.90}                      & \textcolor{gray_lc2}{61.78}                  & \multicolumn{1}{r}{\textcolor{gray_lc2}{74.01}}          \\
\rowcolor[HTML]{a9d18e} 
\textbf{\textsc{DDESeg (Ours)}}                                           & Trans.                                                 & HTSAT                                                & 103.84M                                                 & \textbf{74.38}            & \textbf{67.55}         & \textbf{81.20}          & \textbf{74.30}             & \textbf{67.64}         & \textbf{80.96}         & \textbf{68.39}            & \textbf{62.11}         & \multicolumn{1}{r}{\textbf{74.67}} \\ \bottomrule
\end{tabular}
\vspace{-2.0em}
\end{table*}

\section{Experiments}
\noindent\textbf{Architecture Details.}
Our framework includes four feature fusion modules, each containing 3, 6, 40, and 3 fusion blocks, respectively. 
The feature fusion module is modified based on MiT \cite{xie2021segformer}.
For audio feature extraction, we adopt HTSAT \cite{chen2022hts}, a state-of-the-art audio classification model pre-trained on the large-scale AudioSet dataset \cite{gemmeke2017audio}, which provides robust audio representations across diverse acoustic environments.

\noindent\textbf{Metrics}. We employ the Jaccard Index ($\mathcal{J}$) and F-score ($\mathcal{F}_{\beta}$) to evaluate model performance, following the setup in \cite{chen2024unraveling}, with $\beta$ set to 0.3.

\noindent\textbf{Datasets}. We evaluate our approach on three benchmark datasets for audio-visual segmentation:
\begin{itemize}
    \item \textbf{AVS-Object} \cite{zhou2022audio} includes two subsets: Single Sound Source (S4) with 4,932 five-second video clips and Multi Sound Sources (MS3) with 424 five-second clips.
    For each video, the last frame of each second is paired with the corresponding audio signal as input. 
    Ground truth is a binary mask, where 1 indicates the sounding region.
    \item \textbf{AVS-Semantic} \cite{zhou2024audio} comprises 12,356 video clips across 71 categories, each lasting 5s or 10s.
    A semantic mask is provided as ground truth for each audio-visual pair, identifying both the sounding region and the audio event type.
    \item \textbf{VPO} \cite{chen2024unraveling} consists of audio-visual pairs created from single images in COCO \cite{lin2014microsoft} combined with three-second audio signals from VGGSound \cite{chen2020vggsound}, with semantic masks provided as ground truth. It covers 21 categories and includes three subsets: VPO-SS (12,202 pairs with a single sound source), VPO-MS (9,817 pairs with multiple sound sources), and VPO-MSMI (12,855 pairs with multiple silent and audible objects of the same category).
\end{itemize}

\subsection{Quantitative Results}
\label{sec:qualitative_ana}
\noindent\textbf{Performance on AVS-Object and AVS-Semantic}.
Table \ref{tab:avss} compares our method with previous state-of-the-art approaches on AVS-Object and AVS-Semantic using transformer-based visual backbones. 
Our method shows substantial improvements across all evaluation metrics, achieving increases in  $\mathcal{J}$\&$\mathcal{F}_\beta$ compared to the second-best method: from 90.5\% to 94.2\% on AVS-Object-S4, 74.0\% to 77.9\% on AVS-Object-MS3, and 61.5\% to 67.9\% on AVS-Semantic.
These gains highlight our method’s capacity for precise sound localization, even in complex multi-source semantic segmentation scenarios.
Note that our visual backbone has a comparable parameter (85M) to Swin-Base \cite{liu2021swintransformerhierarchicalvision} (88M) and PVT-V2-B5 \cite{wang2022pvt} (82M). 
To further validate the effectiveness of our approach, we also adopt PVT-V2-B5 as the visual encoder, which achieves consistent superior performance across all evaluation metrics.

Additionally, we replace VGGish \cite{hershey2017cnnarchitectureslargescaleaudio} with the state-of-the-art audio classification model HTSAT \cite{chen2022hts} as the audio backbone, leading to performance improvements of 1.5\% in $\mathcal{J}$ and 0.7\% in $\mathcal{F}_\beta$.
This indicates that enhanced audio representations can positively impact sound localization. 

\noindent\textbf{Performance on VPO}.
We evaluate recent state-of-the-art (SOTA) methods on the VPO dataset equipped with synthesized stereo audios.
As illustrated in Table \ref{tab:vpo}, DDESeg consistently outperforms all existing methods.
Since the code of CPM \cite{chen2024cpm} is unavailable, we report its results at the original image resolution, denoted as CPM* in Table \ref{tab:vpo}.
Despite being trained on lower-resolution images (224$\times$224 pixels), our model still outperforms CPM.
Moreover, DDESeg delivers significant performance gains over other methods with comparable parameter counts and identical input resolution. 
Specifically, compared to the previous best method AVSStone \cite{ma2024stepping}, DDESeg achieves 6\%, 10.04\%, and 4.9\% higher $\mathcal{J}$\&$\mathcal{F}_\beta$ scores on the SS, MS, and MSMI sub-datasets, respectively. 
These substantial improvements on the VPO datasets underscore the effectiveness of our audio enhancement method.
 
\subsection{Qualitative Results}
\label{sec:qualitative_ana}
Fig. \ref{fig:com_avss} presents qualitative comparisons between our method and recent state-of-the-art approaches on AVS-Semantic.
In challenging multi-source scenarios, our method accurately segments multiple simultaneously sounding objects, whereas other methods suffer from segmentation omissions. 
Notably, AVSStone fails to segment the \texttt{`drum'} in the first frame and misclassifies the \texttt{`cello'} as a  \texttt{`marimba'} in the second frame. 
These results highlight the effectiveness of our approach in attaining high-quality audio semantic representations. 
We posit that deriving clearer sound semantics not only enhances localization accuracy but also ensures more reliable sound class identification. 
More visualizations are provided in the \textit{Supplementary Material}.

\subsection{Further Analysis}
\label{sec:ablation_ana}
\noindent\textbf{Ablation Study.}
For detailed analysis, we conduct a series of ablation studies on AVS-Objects-MS3 and AVS-Semantic, as shown in Table \ref{tab:ablation_Com}.
We first establish a baseline performance without DDM and DEM. Building on this baseline, the integration of audio representation derivation (Derr.) has a significant improvement, improving the $\mathcal{J}$\&$\mathcal{F}_\beta$ score by 4.6\% on AVS-Semantic, validating the effectiveness of our representation derivation design.
Adding discriminative semantic enhancement further boosts performance, providing an additional 2.95\% improvement in $\mathcal{J}$\&$\mathcal{F}_\beta$ on AVS-Semantic. 
On this basis, incorporating DEM yields a further 2.33\% increase in the $\mathcal{J}$ score and a 2.27\% increase in the $\mathcal{F}_\beta$ score. 
This confirms that our dynamic elimination design effectively reduces interference from non-visually matching audio representations.

\begin{figure}[t]
\begin{center}
\includegraphics[width=0.90\linewidth]{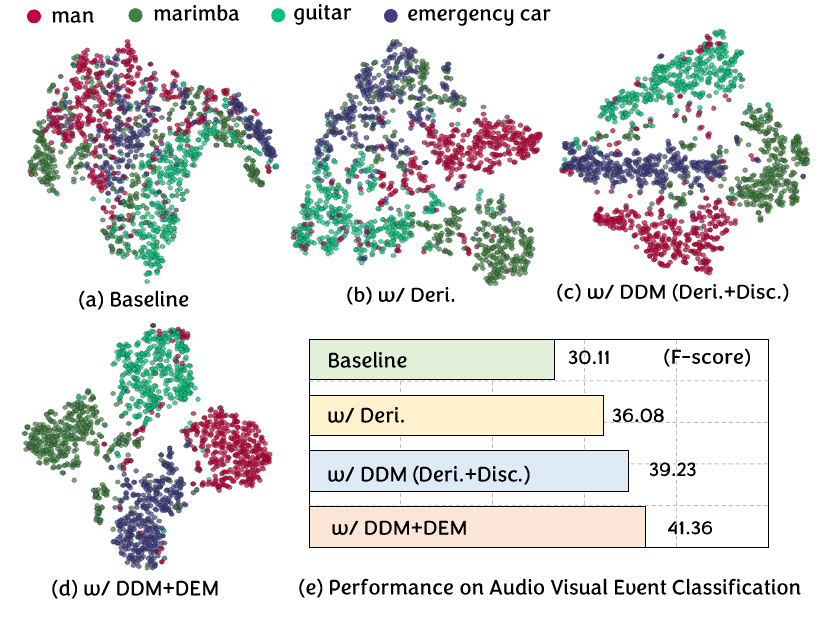}
\end{center}
\vspace{-2em}
\caption{Qualitative and quantitative comparisons of audio semantic quality. (a)-(c) show t-SNE visualizations of audio features under different settings. (d) presents the performance of employing audio-visual interacted semantics from different settings in audio-visual event classification. See \S \ref{sec:ablation_ana} for analysis details.}
\vspace{-0.5em}
\label{fig:aud_vis}
\end{figure}

\begin{table}[t]
\centering
\footnotesize
\setlength{\tabcolsep}{0.7mm}
\renewcommand\arraystretch{0.9}
\caption{Ablation study (see details in \S \ref{sec:ablation_ana}) on AVS datasets.}
\vspace{-1.0em}
\label{tab:ablation_Com}
\begin{tabular}{l|ccc|ccc}
\toprule
\multirow{2}{*}{Method} & \multicolumn{3}{c|}{\textit{AVS-Objects-MS3}}                                        & \multicolumn{3}{c}{\textit{AVS-Semantic}}                                         \\ \cmidrule{2-7} 
                        & \multicolumn{1}{c}{$\mathcal{J}\&\mathcal{F}_\beta \uparrow$}  & \multicolumn{1}{c}{$\mathcal{J} \uparrow$} & \multicolumn{1}{c|}{{$\mathcal{F}_\beta \uparrow$}  } & \multicolumn{1}{c}{$\mathcal{J}\&\mathcal{F}_\beta \uparrow$}  & \multicolumn{1}{c}{$\mathcal{J} \uparrow$} & \multicolumn{1}{c}{{$\mathcal{F}_\beta \uparrow$}} \\ \midrule
Baseline                & 65.81                     & 60.03                  & 71.59                  & 58.00                    & 52.69                 & 63.31                 \\
Deri.              & 71.51                     & 65.20                  & 77.81                  & 62.60                    & 57.14                 & 68.06                 \\
DEM (Deri.+Disc.)  & 73.40                     & 67.50                  & 79.30                  & 65.55                    & 61.07                 & 70.03                 \\
\rowcolor[HTML]{a9d18e} 
DEM+DDM (Ours)          & \textbf{75.57}            & \textbf{70.10}         & \textbf{81.03}         & \textbf{67.85}           & \textbf{63.40}        & \textbf{72.30}        \\ \bottomrule
\end{tabular}
\vspace{-2.0em}
\end{table}
\noindent\textbf{Quality of Generated Audio Representations.}
Fig. \ref{fig:aud_vis} presents the qualitative and quantitative evaluations of audio representation quality across different settings.
A comparison between (a) and (b) shows a distinct clustering trend in audio features with the introduction of audio representation derivation (\textit{w}/ Deri.). 
This clustering effect is further strengthened when both audio representation derivation and discriminative semantic enhancement are applied.
Notably, the intra-class distribution in (b) reveals sub-clustering patterns, as observed in classes such as \texttt{`marimba'} and \texttt{`emergency car'}. 
These variations are effectively mitigated in (c), where the dynamic elimination module (\textit{w}/ DDM+DEM) results in more pronounced class-wise clustering, underscoring the substantial improvements our modules bring to audio quality.

To quantitatively validate our semantic representations, we conducted linear probing experiments by assessing their performance in audio-visual event classification. 
Using features from each model variant as input to a linear classifier, we observe that the \textit{w}/ DDM+DEM configuration achieved an 11.25\% F-score improvement over the baseline. 
This notable performance increase in linear probing provides further evidence that our approach generates high-quality audio-visual representations.

\begin{table}[]
\caption{Further study of elimination designs (\S \ref{sec:ablation_ana}). FC: Fully Connected layer, CA: Cross-Attention Layer, SK: Soft-Kmeans, and GS: Gumbel Softmax.}
\vspace{-0.5em}
\label{tab:elimination_ab}
\centering
\footnotesize
\setlength{\tabcolsep}{0.7mm}
\renewcommand\arraystretch{0.9}
\begin{tabular}{r|l|ccc|ccc}
\toprule
\multicolumn{1}{c|}{\multirow{2}{*}{}} & \multirow{2}{*}{Scheme}                         & \multicolumn{3}{c|}{\textit{AVS-Objects-MS3}}                                      & \multicolumn{3}{c}{\textit{AVS-Semantic}}                                         \\ \cmidrule{3-8} 
\multicolumn{1}{c|}{}                  &                                                 & \multicolumn{1}{c}{$\mathcal{J}\&\mathcal{F}_\beta \uparrow$} & \multicolumn{1}{c}{$\mathcal{J} \uparrow$} & \multicolumn{1}{c|}{$\mathcal{F}_\beta \uparrow$} & \multicolumn{1}{c}{$\mathcal{J}\&\mathcal{F}_\beta \uparrow$} & \multicolumn{1}{c}{$\mathcal{J} \uparrow$} & \multicolumn{1}{c}{$\mathcal{F}_\beta \uparrow$} \\ \midrule
\ding{172}                                      & wo/Elimination                                  & 73.4                     & 67.5                  & 79.3                   & 65.6                     & 61.1                  & 70.0                  \\
\ding{173}                                       & FC+Sigmoid                                      & 72.1                     & 66.7                  & 77.4                   & 64.0                     & 59.8                  & 68.2                  \\
\ding{174}                                      & CA+FC+Sigmoid                       & 74.4                     & 68.4                  & 80.3                   & 65.9                     & 61.8                  & 70.0                  \\
\ding{175}                                         & SK+CA+FC+Sigmoid   & 74.8                     & 69.0                  & 80.6                   & 66.5                     & 62.1                  & 70.9                  \\
\rowcolor[HTML]{a9d18e} 
\ding{176}                                      & GS+CA+FC+Sigmoid & \textbf{77.9}            & \textbf{72.3}         & \textbf{83.4}          & \textbf{67.9}            & \textbf{63.4}         & \textbf{72.3}         \\ \bottomrule
\end{tabular}
\vspace{-2.0em}
\end{table}

\noindent\textbf{Analysis of Elimination Designs.}
We systematically investigate various elimination designs to optimize audio-visual correlation learning, with results presented in Table \ref{tab:elimination_ab}. This analysis explores five progressive architectural variants:
\begin{itemize}
\item \textit{Scheme \ding{172}} (Baseline): Directly interacts derived audio semantic representations with visual feature maps through a feature fusion module. 
\item \textit{Scheme \ding{173}}: Incorporates the FC-Sigmoid filtering mechanism, which results in degraded performance.
\item \textit{Scheme \ding{174}}: Enhances audio-visual correlation by applying cross-attention prior to filtering, yielding a significant improvement of 0.3\% in $\mathcal{J}$\&$\mathcal{F}_\beta$ compared to Scheme \ding{172}.
\item \textit{Scheme \ding{175}}: Adds Soft-Kmeans to attain more compact visual semantic representations, leading to a 0.6\% improvement in $\mathcal{J}$\&$\mathcal{F}_\beta$ over Scheme \ding{174}. 
\item \textit{Scheme \ding{176}}: Replaces Soft-Kmeans with Gumbel Softmax for enhanced visual semantic compactness, achieving a 1.4\% improvement in $\mathcal{J}$\&$\mathcal{F}_\beta$ over scheme \ding{175}.
\end{itemize}
As the results suggested, Scheme \ding{176} attains the best performance, demonstrating that visual semantic compaction is essential for robust audio-visual correlation learning.

\section{Conclusion}
\label{Sec:conclusion}
In this paper, we propose an effective framework for audio-visual segmentation that addresses challenges posed by complex audio characteristics. Central to our approach are two key modules: the Dynamic Derivation Module (DDM), which generates multiple audio representations with distinct semantics, and the Dynamic Elimination Module (DEM), which filters out non-visually matching audio cues.
Within the DDM, the derivation module addresses issues in feature fusion due to the overlapping nature of audio, while the discriminative semantic enhancement component mitigates the audio-visual matching difficulty caused by great intra-class variations. 
Leveraging DEM, our framework achieves precise audio-visual alignment.
Extensive experiments on three benchmark datasets demonstrate that our approach significantly advances the state-of-the-art, showing particular strength in challenging multi-source scenarios.

\vspace{0.5em}
\noindent\textbf{Acknowledgements.} This work is supported by ARC-Discovery (DP220100800 to XY) and ARC-DECRA  (DE230100477 to XY). Chen Liu is funded by the China Scholarship Council and CSIRO top-up (50092128). 
{
    \small
    \bibliographystyle{ieeenat_fullname}
    \bibliography{main}

\begin{thebibliography}{56}
\providecommand{\natexlab}[1]{#1}
\providecommand{\url}[1]{\texttt{#1}}
\expandafter\ifx\csname urlstyle\endcsname\relax
  \providecommand{\doi}[1]{doi: #1}\else
  \providecommand{\doi}{doi: \begingroup \urlstyle{rm}\Url}\fi

\bibitem[Arandjelovic and Zisserman(2018)]{arandjelovic2018objects}
Relja Arandjelovic and Andrew Zisserman.
\newblock Objects that sound.
\newblock In \emph{Proceedings of the European conference on computer vision (ECCV)}, pages 435--451, 2018.

\bibitem[Belkin and Niyogi(2003)]{belkin2003laplacian}
Mikhail Belkin and Partha Niyogi.
\newblock Laplacian eigenmaps for dimensionality reduction and data representation.
\newblock \emph{Neural computation}, 15\penalty0 (6):\penalty0 1373--1396, 2003.

\bibitem[Chen et~al.(2020)Chen, Xie, Vedaldi, and Zisserman]{chen2020vggsound}
Honglie Chen, Weidi Xie, Andrea Vedaldi, and Andrew Zisserman.
\newblock Vggsound: A large-scale audio-visual dataset.
\newblock In \emph{ICASSP 2020-2020 IEEE International Conference on Acoustics, Speech and Signal Processing (ICASSP)}, pages 721--725. IEEE, 2020.

\bibitem[Chen et~al.(2022)Chen, Du, Zhu, Ma, Berg-Kirkpatrick, and Dubnov]{chen2022hts}
Ke Chen, Xingjian Du, Bilei Zhu, Zejun Ma, Taylor Berg-Kirkpatrick, and Shlomo Dubnov.
\newblock Hts-at: A hierarchical token-semantic audio transformer for sound classification and detection.
\newblock In \emph{ICASSP 2022-2022 IEEE International Conference on Acoustics, Speech and Signal Processing (ICASSP)}, pages 646--650. IEEE, 2022.

\bibitem[Chen et~al.(2024{\natexlab{a}})Chen, Liu, Wang, Liu, Wang, Frazer, and Carneiro]{chen2024unraveling}
Yuanhong Chen, Yuyuan Liu, Hu Wang, Fengbei Liu, Chong Wang, Helen Frazer, and Gustavo Carneiro.
\newblock Unraveling instance associations: A closer look for audio-visual segmentation.
\newblock In \emph{Proceedings of the IEEE/CVF Conference on Computer Vision and Pattern Recognition}, pages 26497--26507, 2024{\natexlab{a}}.

\bibitem[Chen et~al.(2024{\natexlab{b}})Chen, Wang, Liu, Wang, and Carneiro]{chen2024cpm}
Yuanhong Chen, Chong Wang, Yuyuan Liu, Hu Wang, and Gustavo Carneiro.
\newblock Cpm: Class-conditional prompting machine for audio-visual segmentation.
\newblock \emph{arXiv preprint arXiv:2407.05358}, 2024{\natexlab{b}}.

\bibitem[Cheng et~al.(2021)Cheng, Schwing, and Kirillov]{cheng2021maskformer}
Bowen Cheng, Alexander~G. Schwing, and Alexander Kirillov.
\newblock Per-pixel classification is not all you need for semantic segmentation.
\newblock 2021.

\bibitem[Cheng et~al.(2022)Cheng, Misra, Schwing, Kirillov, and Girdhar]{cheng2021mask2former}
Bowen Cheng, Ishan Misra, Alexander~G. Schwing, Alexander Kirillov, and Rohit Girdhar.
\newblock Masked-attention mask transformer for universal image segmentation.
\newblock 2022.

\bibitem[Chng et~al.(2024)Chng, Zheng, Han, Qiu, and Huang]{chng2024mask}
Yong~Xien Chng, Henry Zheng, Yizeng Han, Xuchong Qiu, and Gao Huang.
\newblock Mask grounding for referring image segmentation.
\newblock In \emph{Proceedings of the IEEE/CVF Conference on Computer Vision and Pattern Recognition}, pages 26573--26583, 2024.

\bibitem[Ding et~al.(2021)Ding, Liu, Wang, and Jiang]{ding2021vision}
Henghui Ding, Chang Liu, Suchen Wang, and Xudong Jiang.
\newblock Vision-language transformer and query generation for referring segmentation.
\newblock In \emph{Proceedings of the IEEE/CVF International Conference on Computer Vision}, pages 16321--16330, 2021.

\bibitem[Ding et~al.(2023)Ding, Liu, He, Jiang, and Loy]{ding2023mevis}
Henghui Ding, Chang Liu, Shuting He, Xudong Jiang, and Chen~Change Loy.
\newblock Mevis: A large-scale benchmark for video segmentation with motion expressions.
\newblock In \emph{Proceedings of the IEEE/CVF International Conference on Computer Vision}, pages 2694--2703, 2023.

\bibitem[Gao et~al.(2024)Gao, Chen, Chen, Wang, and Lu]{gao2024avsegformer}
Shengyi Gao, Zhe Chen, Guo Chen, Wenhai Wang, and Tong Lu.
\newblock Avsegformer: Audio-visual segmentation with transformer.
\newblock In \emph{Proceedings of the AAAI Conference on Artificial Intelligence}, pages 12155--12163, 2024.

\bibitem[Gemmeke et~al.(2017)Gemmeke, Ellis, Freedman, Jansen, Lawrence, Moore, Plakal, and Ritter]{gemmeke2017audio}
Jort~F Gemmeke, Daniel~PW Ellis, Dylan Freedman, Aren Jansen, Wade Lawrence, R~Channing Moore, Manoj Plakal, and Marvin Ritter.
\newblock Audio set: An ontology and human-labeled dataset for audio events.
\newblock In \emph{2017 IEEE international conference on acoustics, speech and signal processing (ICASSP)}, pages 776--780. IEEE, 2017.

\bibitem[Hao et~al.(2024)Hao, Mao, He, Han, Dai, and Zhong]{hao2024improving}
Dawei Hao, Yuxin Mao, Bowen He, Xiaodong Han, Yuchao Dai, and Yiran Zhong.
\newblock Improving audio-visual segmentation with bidirectional generation.
\newblock In \emph{Proceedings of the AAAI Conference on Artificial Intelligence}, pages 2067--2075, 2024.

\bibitem[Hershey et~al.(2017)Hershey, Chaudhuri, Ellis, Gemmeke, Jansen, Moore, Plakal, Platt, Saurous, Seybold, Slaney, Weiss, and Wilson]{hershey2017cnnarchitectureslargescaleaudio}
Shawn Hershey, Sourish Chaudhuri, Daniel P.~W. Ellis, Jort~F. Gemmeke, Aren Jansen, R.~Channing Moore, Manoj Plakal, Devin Platt, Rif~A. Saurous, Bryan Seybold, Malcolm Slaney, Ron~J. Weiss, and Kevin Wilson.
\newblock Cnn architectures for large-scale audio classification, 2017.

\bibitem[Hu et~al.(2023)Hu, Wang, Shao, Xie, Li, Han, and Luo]{hu2023beyond}
Yutao Hu, Qixiong Wang, Wenqi Shao, Enze Xie, Zhenguo Li, Jungong Han, and Ping Luo.
\newblock Beyond one-to-one: Rethinking the referring image segmentation.
\newblock In \emph{Proceedings of the IEEE/CVF International Conference on Computer Vision}, pages 4067--4077, 2023.

\bibitem[Huang et~al.(2023)Huang, Li, Wang, Zhu, Dai, Han, Rong, and Liu]{huang2023discovering}
Shaofei Huang, Han Li, Yuqing Wang, Hongji Zhu, Jiao Dai, Jizhong Han, Wenge Rong, and Si Liu.
\newblock Discovering sounding objects by audio queries for audio visual segmentation.
\newblock \emph{arXiv preprint arXiv:2309.09501}, 2023.

\bibitem[Hui et~al.(2023)Hui, Liu, Ding, Huang, Li, Wang, Liu, and Han]{hui2023language}
Tianrui Hui, Si Liu, Zihan Ding, Shaofei Huang, Guanbin Li, Wenguan Wang, Luoqi Liu, and Jizhong Han.
\newblock Language-aware spatial-temporal collaboration for referring video segmentation.
\newblock \emph{IEEE Transactions on Pattern Analysis and Machine Intelligence}, 45\penalty0 (7):\penalty0 8646--8659, 2023.

\bibitem[Kim et~al.(2022)Kim, Kim, Lan, Zeng, and Kwak]{kim2022restr}
Namyup Kim, Dongwon Kim, Cuiling Lan, Wenjun Zeng, and Suha Kwak.
\newblock Restr: Convolution-free referring image segmentation using transformers.
\newblock In \emph{Proceedings of the IEEE/CVF Conference on Computer Vision and Pattern Recognition}, pages 18145--18154, 2022.

\bibitem[Li et~al.(2022)Li, Wei, Tian, Xu, Wen, and Hu]{li2022learning}
Guangyao Li, Yake Wei, Yapeng Tian, Chenliang Xu, Ji-Rong Wen, and Di Hu.
\newblock Learning to answer questions in dynamic audio-visual scenarios.
\newblock In \emph{Proceedings of the IEEE/CVF Conference on Computer Vision and Pattern Recognition}, pages 19108--19118, 2022.

\bibitem[Li et~al.(2023{\natexlab{a}})Li, Hou, and Hu]{li2023progressive}
Guangyao Li, Wenxuan Hou, and Di Hu.
\newblock Progressive spatio-temporal perception for audio-visual question answering.
\newblock In \emph{Proceedings of the 31st ACM International Conference on Multimedia}, pages 7808--7816, 2023{\natexlab{a}}.

\bibitem[Li et~al.(2023{\natexlab{b}})Li, Yang, Chen, Yang, and Xiao]{li2023catr}
Kexin Li, Zongxin Yang, Lei Chen, Yi Yang, and Jun Xiao.
\newblock Catr: Combinatorial-dependence audio-queried transformer for audio-visual video segmentation.
\newblock In \emph{Proceedings of the 31st ACM International Conference on Multimedia}, pages 1485--1494, 2023{\natexlab{b}}.

\bibitem[Li et~al.(2019)Li, Sun, Meng, Liang, Wu, and Li]{li2019dice}
Xiaoya Li, Xiaofei Sun, Yuxian Meng, Junjun Liang, Fei Wu, and Jiwei Li.
\newblock Dice loss for data-imbalanced nlp tasks.
\newblock \emph{arXiv preprint arXiv:1911.02855}, 2019.

\bibitem[Li et~al.(2024{\natexlab{a}})Li, Wang, Xu, Peng, Singh, Lu, and Raj]{li2024qdformer}
Xiang Li, Jinglu Wang, Xiaohao Xu, Xiulian Peng, Rita Singh, Yan Lu, and Bhiksha Raj.
\newblock Qdformer: Towards robust audiovisual segmentation in complex environments with quantization-based semantic decomposition.
\newblock In \emph{Proceedings of the IEEE/CVF Conference on Computer Vision and Pattern Recognition}, pages 3402--3413, 2024{\natexlab{a}}.

\bibitem[Li et~al.(2024{\natexlab{b}})Li, Guo, Zhou, Zhang, and Wang]{li2024object}
Zhangbin Li, Dan Guo, Jinxing Zhou, Jing Zhang, and Meng Wang.
\newblock Object-aware adaptive-positivity learning for audio-visual question answering.
\newblock In \emph{Proceedings of the AAAI Conference on Artificial Intelligence}, pages 3306--3314, 2024{\natexlab{b}}.

\bibitem[Lin et~al.(2014)Lin, Maire, Belongie, Hays, Perona, Ramanan, Doll{\'a}r, and Zitnick]{lin2014microsoft}
Tsung-Yi Lin, Michael Maire, Serge Belongie, James Hays, Pietro Perona, Deva Ramanan, Piotr Doll{\'a}r, and C~Lawrence Zitnick.
\newblock Microsoft coco: Common objects in context.
\newblock In \emph{Computer Vision--ECCV 2014: 13th European Conference, Zurich, Switzerland, September 6-12, 2014, Proceedings, Part V 13}, pages 740--755. Springer, 2014.

\bibitem[Liu et~al.(2023{\natexlab{a}})Liu, Li, Qi, Zhang, Li, Wang, and Yu]{liu2023audio}
Chen Liu, Peike~Patrick Li, Xingqun Qi, Hu Zhang, Lincheng Li, Dadong Wang, and Xin Yu.
\newblock Audio-visual segmentation by exploring cross-modal mutual semantics.
\newblock In \emph{Proceedings of the 31st ACM International Conference on Multimedia}, pages 7590--7598, 2023{\natexlab{a}}.

\bibitem[Liu et~al.(2024{\natexlab{a}})Liu, Li, Zhang, Li, Huang, Wang, and Yu]{liu2024bavs}
Chen Liu, Peike Li, Hu Zhang, Lincheng Li, Zi Huang, Dadong Wang, and Xin Yu.
\newblock Bavs: bootstrapping audio-visual segmentation by integrating foundation knowledge.
\newblock \emph{IEEE Transactions on Multimedia}, 2024{\natexlab{a}}.

\bibitem[Liu et~al.(2024{\natexlab{b}})Liu, Li, and Ding]{liu2024referring}
Chang Liu, Xiangtai Li, and Henghui Ding.
\newblock Referring image editing: Object-level image editing via referring expressions.
\newblock In \emph{Proceedings of the IEEE/CVF Conference on Computer Vision and Pattern Recognition}, pages 13128--13138, 2024{\natexlab{b}}.

\bibitem[Liu et~al.(2023{\natexlab{b}})Liu, Ju, Ma, Wang, Wang, and Zhang]{liu2023audio_a}
Jinxiang Liu, Chen Ju, Chaofan Ma, Yanfeng Wang, Yu Wang, and Ya Zhang.
\newblock Audio-aware query-enhanced transformer for audio-visual segmentation.
\newblock \emph{arXiv preprint arXiv:2307.13236}, 2023{\natexlab{b}}.

\bibitem[Liu et~al.(2024{\natexlab{c}})Liu, Wang, Ju, Ma, Zhang, and Xie]{liu2024annotation}
Jinxiang Liu, Yu Wang, Chen Ju, Chaofan Ma, Ya Zhang, and Weidi Xie.
\newblock Annotation-free audio-visual segmentation.
\newblock In \emph{Proceedings of the IEEE/CVF Winter Conference on Applications of Computer Vision}, pages 5604--5614, 2024{\natexlab{c}}.

\bibitem[Liu et~al.(2021)Liu, Lin, Cao, Hu, Wei, Zhang, Lin, and Guo]{liu2021swintransformerhierarchicalvision}
Ze Liu, Yutong Lin, Yue Cao, Han Hu, Yixuan Wei, Zheng Zhang, Stephen Lin, and Baining Guo.
\newblock Swin transformer: Hierarchical vision transformer using shifted windows, 2021.

\bibitem[Luo et~al.(2024)Luo, Xiao, Liu, Li, Wang, Tang, Li, and Yang]{luo2024soc}
Zhuoyan Luo, Yicheng Xiao, Yong Liu, Shuyan Li, Yitong Wang, Yansong Tang, Xiu Li, and Yujiu Yang.
\newblock Soc: Semantic-assisted object cluster for referring video object segmentation.
\newblock \emph{Advances in Neural Information Processing Systems}, 36, 2024.

\bibitem[Ma et~al.(2024)Ma, Sun, Wang, and Hu]{ma2024stepping}
Juncheng Ma, Peiwen Sun, Yaoting Wang, and Di Hu.
\newblock Stepping stones: A progressive training strategy for audio-visual semantic segmentation.
\newblock \emph{arXiv preprint arXiv:2407.11820}, 2024.

\bibitem[Mao et~al.(2023{\natexlab{a}})Mao, Zhang, Xiang, Lv, Zhong, and Dai]{mao2023contrastive}
Yuxin Mao, Jing Zhang, Mochu Xiang, Yunqiu Lv, Yiran Zhong, and Yuchao Dai.
\newblock Contrastive conditional latent diffusion for audio-visual segmentation.
\newblock \emph{arXiv preprint arXiv:2307.16579}, 2023{\natexlab{a}}.

\bibitem[Mao et~al.(2023{\natexlab{b}})Mao, Zhang, Xiang, Zhong, and Dai]{mao2023multimodal}
Yuxin Mao, Jing Zhang, Mochu Xiang, Yiran Zhong, and Yuchao Dai.
\newblock Multimodal variational auto-encoder based audio-visual segmentation.
\newblock In \emph{Proceedings of the IEEE/CVF International Conference on Computer Vision}, pages 954--965, 2023{\natexlab{b}}.

\bibitem[Miao et~al.(2023)Miao, Bennamoun, Gao, and Mian]{miao2023spectrum}
Bo Miao, Mohammed Bennamoun, Yongsheng Gao, and Ajmal Mian.
\newblock Spectrum-guided multi-granularity referring video object segmentation.
\newblock In \emph{Proceedings of the IEEE/CVF International Conference on Computer Vision}, pages 920--930, 2023.

\bibitem[Owens and Efros(2018)]{owens2018audio}
Andrew Owens and Alexei~A Efros.
\newblock Audio-visual scene analysis with self-supervised multisensory features.
\newblock In \emph{Proceedings of the European conference on computer vision (ECCV)}, pages 631--648, 2018.

\bibitem[Sun et~al.(2024)Sun, Zhang, and Hu]{sun2024unveiling}
Peiwen Sun, Honggang Zhang, and Di Hu.
\newblock Unveiling and mitigating bias in audio visual segmentation.
\newblock \emph{arXiv preprint arXiv:2407.16638}, 2024.

\bibitem[Vaswani(2017)]{vaswani2017attention}
A Vaswani.
\newblock Attention is all you need.
\newblock \emph{Advances in Neural Information Processing Systems}, 2017.

\bibitem[Wang et~al.(2022{\natexlab{a}})Wang, Xie, Li, Fan, Song, Liang, Lu, Luo, and Shao]{wang2022pvt}
Wenhai Wang, Enze Xie, Xiang Li, Deng-Ping Fan, Kaitao Song, Ding Liang, Tong Lu, Ping Luo, and Ling Shao.
\newblock Pvt v2: Improved baselines with pyramid vision transformer.
\newblock \emph{Computational Visual Media}, 8\penalty0 (3):\penalty0 415--424, 2022{\natexlab{a}}.

\bibitem[Wang et~al.(2024)Wang, Liu, Li, Ding, Hu, and Li]{wang2024prompting}
Yaoting Wang, Weisong Liu, Guangyao Li, Jian Ding, Di Hu, and Xi Li.
\newblock Prompting segmentation with sound is generalizable audio-visual source localizer.
\newblock In \emph{Proceedings of the AAAI Conference on Artificial Intelligence}, pages 5669--5677, 2024.

\bibitem[Wang et~al.(2022{\natexlab{b}})Wang, Lu, Li, Tao, Guo, Gong, and Liu]{wang2022cris}
Zhaoqing Wang, Yu Lu, Qiang Li, Xunqiang Tao, Yandong Guo, Mingming Gong, and Tongliang Liu.
\newblock Cris: Clip-driven referring image segmentation.
\newblock In \emph{Proceedings of the IEEE/CVF conference on computer vision and pattern recognition}, pages 11686--11695, 2022{\natexlab{b}}.

\bibitem[Wu et~al.(2023)Wu, Han, Wang, Dong, Zhang, and Shen]{wu2023referring}
Dongming Wu, Wencheng Han, Tiancai Wang, Xingping Dong, Xiangyu Zhang, and Jianbing Shen.
\newblock Referring multi-object tracking.
\newblock In \emph{Proceedings of the IEEE/CVF conference on computer vision and pattern recognition}, pages 14633--14642, 2023.

\bibitem[Wu et~al.(2024)Wu, Li, Li, Ding, Tong, and Tao]{wu2024towards}
Jianzong Wu, Xiangtai Li, Xia Li, Henghui Ding, Yunhai Tong, and Dacheng Tao.
\newblock Towards robust referring image segmentation.
\newblock \emph{IEEE Transactions on Image Processing}, 2024.

\bibitem[Xie et~al.(2021)Xie, Wang, Yu, Anandkumar, Alvarez, and Luo]{xie2021segformer}
Enze Xie, Wenhai Wang, Zhiding Yu, Anima Anandkumar, Jose~M Alvarez, and Ping Luo.
\newblock Segformer: Simple and efficient design for semantic segmentation with transformers.
\newblock \emph{Advances in neural information processing systems}, 34:\penalty0 12077--12090, 2021.

\bibitem[Xu et~al.(2023)Xu, Huang, Shang, Yuan, Sun, and Liu]{xu2023meta}
Li Xu, Mark~He Huang, Xindi Shang, Zehuan Yuan, Ying Sun, and Jun Liu.
\newblock Meta compositional referring expression segmentation.
\newblock In \emph{Proceedings of the IEEE/CVF Conference on Computer Vision and Pattern Recognition}, pages 19478--19487, 2023.

\bibitem[Yan et~al.(2024)Yan, Zhang, Guo, Chen, Zhang, Li, Qiao, Dong, He, and Gao]{yan2024referred}
Shilin Yan, Renrui Zhang, Ziyu Guo, Wenchao Chen, Wei Zhang, Hongyang Li, Yu Qiao, Hao Dong, Zhongjiang He, and Peng Gao.
\newblock Referred by multi-modality: A unified temporal transformer for video object segmentation.
\newblock In \emph{Proceedings of the AAAI Conference on Artificial Intelligence}, pages 6449--6457, 2024.

\bibitem[Yang et~al.(2024)Yang, Nie, Li, Gao, Guo, Zhen, Yan, and Xiang]{yang2024cooperation}
Qi Yang, Xing Nie, Tong Li, Pengfei Gao, Ying Guo, Cheng Zhen, Pengfei Yan, and Shiming Xiang.
\newblock Cooperation does matter: Exploring multi-order bilateral relations for audio-visual segmentation.
\newblock In \emph{Proceedings of the IEEE/CVF Conference on Computer Vision and Pattern Recognition}, pages 27134--27143, 2024.

\bibitem[Yu et~al.(2016)Yu, Jiang, Wang, Cao, and Huang]{yu2016unitbox}
Jiahui Yu, Yuning Jiang, Zhangyang Wang, Zhimin Cao, and Thomas Huang.
\newblock Unitbox: An advanced object detection network.
\newblock In \emph{Proceedings of the 24th ACM international conference on Multimedia}, pages 516--520, 2016.

\bibitem[Zhao et~al.(2018)Zhao, Gan, Rouditchenko, Vondrick, McDermott, and Torralba]{zhao2018sound}
Hang Zhao, Chuang Gan, Andrew Rouditchenko, Carl Vondrick, Josh McDermott, and Antonio Torralba.
\newblock The sound of pixels.
\newblock In \emph{Proceedings of the European conference on computer vision (ECCV)}, pages 570--586, 2018.

\bibitem[Zhao et~al.(2019)Zhao, Gan, Ma, and Torralba]{zhao2019sound}
Hang Zhao, Chuang Gan, Wei-Chiu Ma, and Antonio Torralba.
\newblock The sound of motions.
\newblock In \emph{Proceedings of the IEEE/CVF International Conference on Computer Vision}, pages 1735--1744, 2019.

\bibitem[Zhou et~al.(2022)Zhou, Wang, Zhang, Sun, Zhang, Birchfield, Guo, Kong, Wang, and Zhong]{zhou2022audio}
Jinxing Zhou, Jianyuan Wang, Jiayi Zhang, Weixuan Sun, Jing Zhang, Stan Birchfield, Dan Guo, Lingpeng Kong, Meng Wang, and Yiran Zhong.
\newblock Audio--visual segmentation.
\newblock In \emph{European Conference on Computer Vision}, pages 386--403. Springer, 2022.

\bibitem[Zhou et~al.(2024)Zhou, Shen, Wang, Zhang, Sun, Zhang, Birchfield, Guo, Kong, Wang, et~al.]{zhou2024audio}
Jinxing Zhou, Xuyang Shen, Jianyuan Wang, Jiayi Zhang, Weixuan Sun, Jing Zhang, Stan Birchfield, Dan Guo, Lingpeng Kong, Meng Wang, et~al.
\newblock Audio-visual segmentation with semantics.
\newblock \emph{International Journal of Computer Vision}, pages 1--21, 2024.

\bibitem[Zou et~al.(2023)Zou, Dou, Yang, Gan, Li, Li, Dai, Behl, Wang, Yuan, et~al.]{zou2023generalized}
Xueyan Zou, Zi-Yi Dou, Jianwei Yang, Zhe Gan, Linjie Li, Chunyuan Li, Xiyang Dai, Harkirat Behl, Jianfeng Wang, Lu Yuan, et~al.
\newblock Generalized decoding for pixel, image, and language.
\newblock In \emph{Proceedings of the IEEE/CVF Conference on Computer Vision and Pattern Recognition}, pages 15116--15127, 2023.

\bibitem[Zou et~al.(2024)Zou, Yang, Zhang, Li, Li, Wang, Wang, Gao, and Lee]{zou2024segment}
Xueyan Zou, Jianwei Yang, Hao Zhang, Feng Li, Linjie Li, Jianfeng Wang, Lijuan Wang, Jianfeng Gao, and Yong~Jae Lee.
\newblock Segment everything everywhere all at once.
\newblock \emph{Advances in Neural Information Processing Systems}, 36, 2024.

\end{thebibliography}
}

\end{document}